\documentclass[twocolumn,showpacs,preprintnumbers,amsmath,amssymb]{revtex4}
\usepackage{graphicx}
\usepackage{dcolumn}
\usepackage{bm}

\begin{document}
\title{The role of multiple soliton and breather interactions in generation
of rogue waves: the mKdV framework}
\author{A.V. Slunyaev$^{1,2}$}
\author{E.N. Pelinovsky$^{1,2}$}

\affiliation{$^1$Institute of Applied Physics, 46 Ulyanova Street, N. Novgorod 603950, Russia \\
$^2$Nizhny Novgorod State Technical University, 24 Minina Street, N. Novgorod 603950, Russia}

\date{\today}

\begin{abstract}
The role of multiple soliton and breather interactions in formation of very high waves is disclosed within the framework of integrable modified Korteweg - de Vries (mKdV) equation. Optimal conditions for the focusing of many solitons are formulated explicitly. Namely, trains of ordered solitons with alternate polarities evolve to huge strongly localized transient waves. The focused wave amplitude is exactly the sum of the focusing soliton heights; the maximum wave inherits the polarity of the fastest soliton in the train. The focusing of several solitary waves or/and breathers may naturally occur in a soliton gas and will lead to rogue-wave-type dynamics; hence it represents a new nonlinear mechanism of rogue wave generation. The discovered scenario depends crucially on the soliton polarities (phases), and thus cannot be taken into account by existing kinetic theories. The performance of the soliton mechanism of rogue wave generation is shown for the example of focusing mKdV equation, when solitons possess 'frozen' phases (polarities), though the approach works in other integrable systems which admit soliton and breather solutions.
\end{abstract}

\maketitle
\emph{Introduction.---}
The generation of unexpectedly large waves from stochastic fields has attracted much interest in many recent studies thanks to recognition of the rogue wave phenomenon in marine and optical realms (see \cite{KharifPelinovsky2003}
and many others).
The modulational instability is the most recognized physical effect capable of generation of very high waves due to the energy transfer from many waves towards an inoculating perturbation of the wavetrain. Most importantly, the nonlinear dynamics alters essentially the statistical properties of stochastic waves, favouring occurrence of very high waves. The accounting for significant deviations from the quasi Gaussian states breaks down the classic assumptions of the wave turbulence theory.
The wave phase averaging becomes inappropriate, thus direct simulations of irregular waves are involved to discover the statistics of high waves.

The strongly nonlinear limit of irregular waves in weakly dispersive media may be treated as a soliton gas which is another intriguing topic of the modern science. Kinetic equations for soliton ensembles were derived in \cite{Zakharov1971}.
It is essential that these equations describe transport of eigenvalues in space but do not concern soliton phases.
Besides general approaches for description of the integrable turbulence considerable understanding may be achieved by virtue of simplified problem statements. In particular, the effect of soliton collisions on statistical moments in integrable Korteweg - de Vries (KdV) and modified KdV (mKdV) equations was considered in \cite{Pelinovskyetal2013,PelinovskyShurgalina2015} through the prism of a two-soliton interaction. These classic equations govern wave dynamics in various important applications. In the course of interaction of two KdV solitons the maximum displacement remains always below the maximum soliton amplitude. Collisions of unipolar mKdV solitons behave similarly to the KdV solitons and never cause larger waves. This point was confirmed in direct numerical simulations within integrable as well as non-integrable equations of the KdV type \cite{DutykhPelinovsky2014,ShurgalinaPelinovsky2016}.
When the conventional definition of a rogue wave is adopted for function of space and time $u(x, t)$ in form
\begin{align}\label{RWDefinition}
AI \equiv \frac{\max_x{|u(x,t)|}}{\max_x{|u(x,t \rightarrow - \infty)|}} > 2 ,
\end{align}
then in the all considered cases of unipolar soliton interactions $AI \le 1$.

Two soliton phases (polarities) are allowed in the mKdV framework due to the isotropic nonlinearity.
In contrast to the  KdV soliton collision,
when two mKdV solitons of opposite polarity interact,
the faster soliton virtually absorbs the smaller soliton rising, and emits it back when overtaken \cite{Slunyaev2001,PelinovskyShurgalina2015}. Surprisingly, the maximum of the transient large wave is given by exact sum of the heights of colliding solitons, and then the attainable amplification is $AI \le 2$.
Hence, the peculiarity of the 'absorb-emit' collision of bipolar solitons yields occurrence of higher waves than could happen in the situation of a unipolar soliton gas. In particular, the occurrence of high waves which are twice higher than the typical soliton height was observed in numerical simulations  \cite{ShurgalinaPelinovsky2016}.

%
%
The competitive roles of solitons and dispersive trains in formation of strongly amplified waves were estimated within the frameworks of the KdV and nonlinear Schr\"{o}dinger (NLS) equations in \cite{Pelinovskyetal2000}.
The analysis of the soliton composition of designed rogue waves in the form of one-humped perturbations revealed a surprising fact: a big wave characterized by $AI > 2$ consists of maximum one soliton. The presence of many solitons in the train which focuses due to the difference in local wave velocities prevents formation of a very high wave.

The crucial distinction from the envelope soliton dynamics in optical fibers, where multiple soliton interactions is a recognized mechanism of rogue wave generation \cite{Mussotetal2009}, should be stressed:
i)~solitons of the mKdV equation preserve their phases 'frozen' (do not change polarity), and ii)~they interact purely elastically and thus do not form giant pulses via fusion or acquiring energy from many smaller solitons.


Thus the role of multiple soliton collisions in spontaneous generation of very high waves has not been clarified so far; it is addressed in this study within the framework of the mKdV equation. In particular, we suggest general conditions when many solitons or/and breathers focus in optimal phase, providing superposition of their partial amplitudes. The process of huge wave formation occurs rapidly and to a large degree unexpectedly, thus conforms with all attributes of the rogue wave phenomenon.

\emph{Generation of rogue waves as a result of multiple soliton collisions.}
In this paper the standard form of the modified Korteweg - de Vries equation (mKdV) with the focusing type of nonlinearity is used
\begin{align}\label{mKdVEquation}
u_t + 6 u^2 u_x + u_{xxx} = 0 ,
\end{align}
where $u(x, t)$ is real. The equation is solvable by means of the Inverse Scattering Transform \cite{Wadati1973} using the fact that the spectrum of the associated scattering problem does not evolve in time. Discrete complex eigenvalues $\{ \lambda \}$ generally appear in quartets and correspond to spatially localized solutions (solitons and breathers), which represent the long-term solution of the Cauchy problem with decaying boundary conditions. A pair of real discrete eigenvalues $\{ \pm \lambda \}$ is responsible for one soliton
\begin{align}\label{mKdVSoliton}
u_s(x,t) = a / \cosh{ \left[{a (x-x_0) -a^3 t} \right]} ,
\end{align}
where $a$ is the soliton amplitude, and $a^2$ is its velocity, related to the eigenvalue as $|a| = 2 |\lambda|$. As usual, the spectrum contains information on neither the initial location of the soliton, $x_0$, nor its phase, which in the case of solution (3) means polarity of the soliton. Two soliton branches exist depending on the sign of real $a$.

The exact $N$-soliton solution to (2) may be obtained, for example, using the Darboux transform \cite{MatveevSalle1991}. It may be represented in form (see details and references in \cite{Slunyaev2001})
\begin{align}\label{NmKdVSoliton}
u_N(x,t) = -i \frac{\partial}{\partial x} \ln{\frac{W(\psi_{1x},\psi_{2x},...,\psi_{Nx})}{W(\psi_{1},\psi_{2},...,\psi_{N})}} , \\ \psi_j = e^{\Theta} + i e^{-\Theta}, \quad \Theta =  \left[ \mu_j(x-x_j) -4 \mu_j^3t \right] +i \theta_j/2 . \nonumber
\end{align}
Here $W(\cdot)$ denote Wronskians for $N$ eigenfunctions $\psi_j$ (in the denominator), and their spatial derivatives (in the numerator).
Each parameter  $\mu_j$, of the seed function  $\psi_j$, $j = 1, ..., N$ coincides with the resulting eigenvalue of the associated scattering problem $\lambda_j$ up to signs of real and imaginary parts, $|\mu_j|^2 = | \lambda_j|^2$. In contrast to  $\lambda_j$, parameters $\mu_j$ in (\ref{NmKdVSoliton}) are strictly definite and hence values $\mu_j$ will be used hereafter for description of the eigenspectrum.
%
%
The parameters of solitons which compose the solution relate to the corresponding eigenvalues as $|a_j| = 2 |\mu_j|$,  where $\mu_j$ are real.

The following three statements which provide the basis for construction of focusing nonlinear wave trains may be proved rigorously. We do not reproduce here the proof, it employs inner symmetries of solution (\ref{NmKdVSoliton}) and properties of determinants with the use of the formal representation
%
$$W(\psi_{1},...,\psi_{N})=
\sum_{\alpha}{(-1)^{n} \psi_{\alpha_1} \partial_x \psi_{\alpha_2}\cdot  ... \cdot \partial_x^{N-1} \psi_{\alpha_N}} 
$$ 
(a similar expression reads for $W(\psi_{1x}, ...,  \psi_{Nx})$). Here the summation is performed along all possible combinations of indices $( \alpha_1,  \alpha_2, ...,  \alpha_N)$ taken from the sequence of natural numbers $(1, 2, ..., N)$;
$n( \alpha_1,  \alpha_2, ...,  \alpha_N)$ is the number of inversions between the indices.


  1. The choice of zero initial coordinates and phases, $x_j = 0$ and $\theta_j = 0$ for all $j = 1, ..., N$, makes point $(x = 0, t = 0)$ a local extremum for function $u_N(x, t)$. Functions $u_N(x, t = 0)$ and $u_N(x = 0, t)$ are symmetric with respect to corresponding variables. Thus the point $(0, 0)$ will be hereafter referred to as the \emph{focusing point}.

  2. The solution in the focusing point is specified by parameters $\mu_j$ in a transparent way,
\begin{align}\label{FocusMaximum}
u_N(0,0)= (-1)^{N-1} \sum_{j=1}^{N} 2 \mu_j ,
\end{align}
which holds true for complex $\mu_j$ as well.

  3. In the case of a one-soliton solution
  signs of $\mu_j$  and $a_j$ agree, $a_1 = 2\mu_1$. When $N > 1$, polarities of the partial solitons depend on combinations of all $\mu_j$. The polarity of soliton number $s$ is specified by the sign of the product
\begin{align}\label{FocusSign}
\text{sgn}{ \left[ \mu_s \prod_{j=1,j\neq s}^{N} \left( \mu_j^2 - \mu_s^2\right) \right] }\,.
\end{align}
If the solitons are sorted in ascending order of values $\mu_j^2$, then the first (slowest) soliton has the same polarity as the sign of corresponding $\mu_1$, the second soliton has polarity opposite to $\mu_2$,
and so on. The desired polarities of all solitons may be set by the choice of signs of $\mu_j$.

\begin{figure}
\centerline{\includegraphics[width=7.5cm]{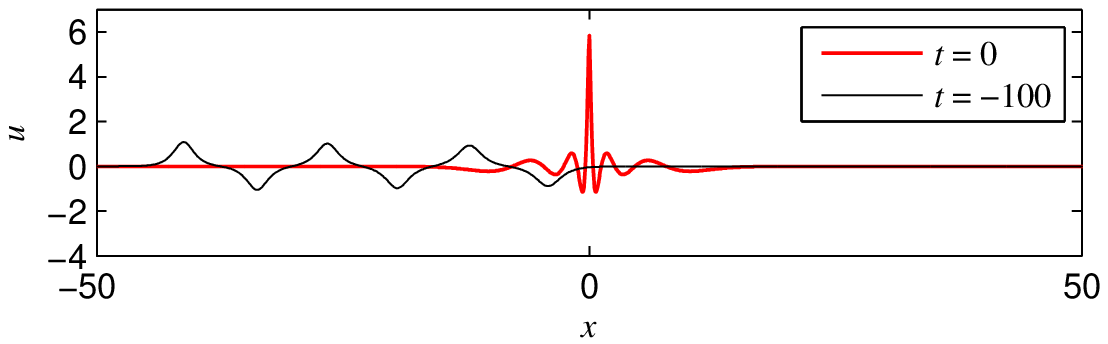}(a)}
\centerline{\includegraphics[width=7.5cm]{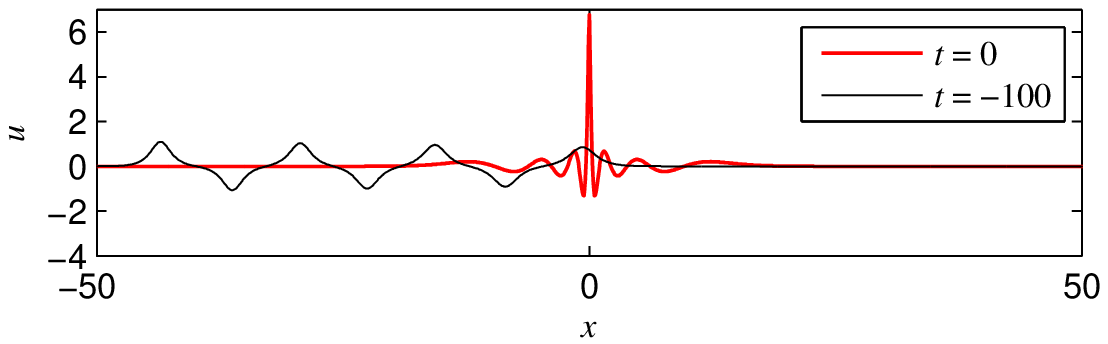}(b)}
\caption{Huge waves caused by focusing of mKdV soliton trains which consist of $6$ (a) and $7$ (b) solitons: instants long before focusing ($t = -100$) and at the focusing ($t=0$). The soliton parameters are (a): $a = \{1,  0.99, 0.98,  0.97, 0.96,  0.95\}$ and (b): $a = \{1,  0.99, 0.98,  0.97, 0.96,  0.95, 0.94\} $.}
\label{fig:1}
\end{figure}

According to (\ref{FocusMaximum}), the most optimal focusing of a soliton train (biggest $|u_N(0, 0)|$) is obviously when all parameters  $\mu_j$ have the same signs.
At instants long before the collision the solitons are located in order of descending velocities,
and then the solitons have alternating polarities according to (\ref{FocusSign}) (thin black lines in Fig.~\ref{fig:1}). It follows from the joint consideration of (\ref{FocusMaximum}) and (\ref{FocusSign}) that  $u_N(0,0)$ inherits polarity of the fastest soliton in the train. The absorb-emit collision of two solitons is a particular case;
 then $a_1 = 2 \mu_1$, $a_2 =  - 2 \mu_2,$ $ \mu_2 >  \mu_1 > 0$, and the focused wave has negative peak. The interaction of unipolar solitons corresponds to alternating signs of $\mu_j$, and then $|u_N(0, 0)|$ never exceeds the height of the largest soliton.

It is easy to see that for given energy (represented by the integral of motion $\int_{-\infty}^{\infty}{u^2 dx}$) the maximum value of $AI$ is achieved when all $\mu_j$ are equal. However, non-degenerative solutions of the inverse scattering problem (when (\ref{NmKdVSoliton}) does not contain singularity) require eigenvalues to be different, and therefore in examples below we consider solitons with different (but close) velocities. We chose the maximum (fastest) soliton of the unit amplitude, $a_1 = 1$ and, correspondingly, unit velocity. All the figures below show solutions of the analytic $N$-soliton formula (\ref{NmKdVSoliton}). In addition, some of the solutions were verified in direct numerical simulations of the mKdV equation;
 In what follows the initial soliton positions and phases  are put equal to zeros, $x_j = 0$, $\theta_j = 0$, $j = 1, ..., N$.

\begin{figure}
\centerline{\includegraphics[width=7.5cm]{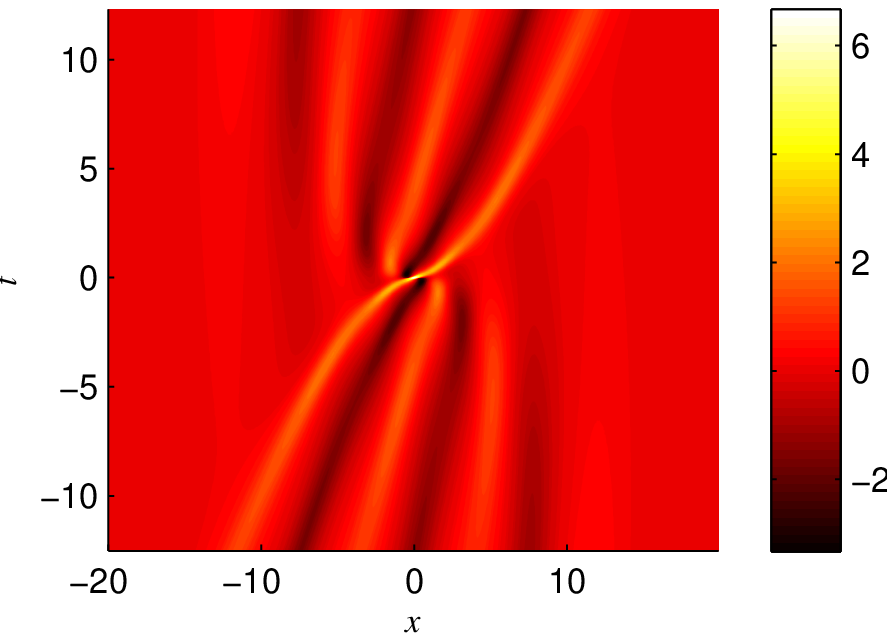}(a)}
\centerline{\includegraphics[width=7.5cm]{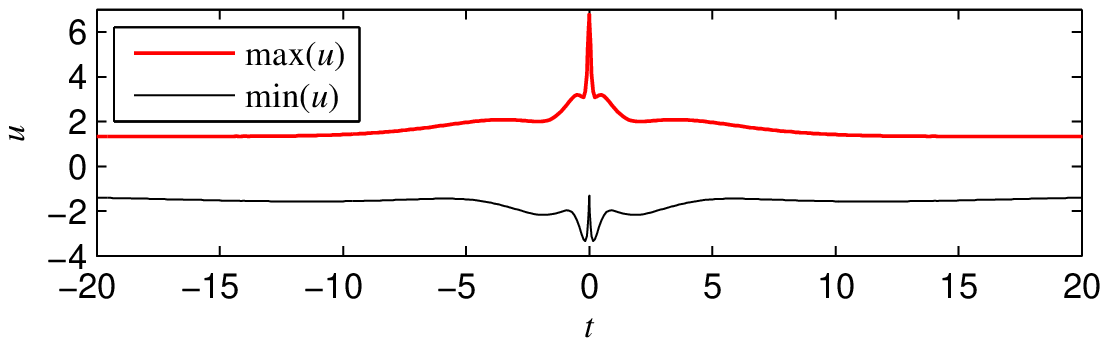}(b)}
\caption{Spatio-temporal plot of $u_N(x,t)$ (a) and corresponding temporal evolution of maximum and minimum values (b) for case $N = 7$ shown in Fig.~\ref{fig:1}b.}
\label{fig:2}
\end{figure}

	Two examples of rogue waves caused by collisions of mKdV solitons are shown in Fig.~\ref{fig:1} for even ($N = 6$) and odd ($N = 7$) number of solitons. The beams of solitons of different polarities before collision are shown by thin lines. The solitons are ordered in velocity so that the faster solitons overtake the slower ones when propagate. The focused waves (thick curves in Fig.~\ref{fig:1}) have complicated shapes and are essentially sign-changing (thus, assumptions implied in study \cite{Pelinovskyetal2000} fail in this case).
They look rather similar in cases of odd and even $N$.
The maxima in Figs.~\ref{fig:1}a,b agree with (\ref{FocusMaximum}), and are equal to $5.85$ and $6.79$ respectively.
Thus, amplification $AI$ may be unlimitedly large if sufficient number of solitons collide.

%
%

	The space-time diagram of the solution, and evolution in time of its maximal and minimal values are shown in Fig.~\ref{fig:2} for the case $N = 7$ displayed in Fig.~\ref{fig:1}b.
The solitons experience strong shifts when collide; the rogue wave lifetime is very short. The fastest soliton  remains the highest wave throughout the collision, it experiences abrupt forward shift. It is clear that the process cannot be interpreted as consequent collisions of soliton pairs, many solitons contribute to the dynamics simultaneously.

\begin{figure}
\centerline{\includegraphics[width=7.5cm]{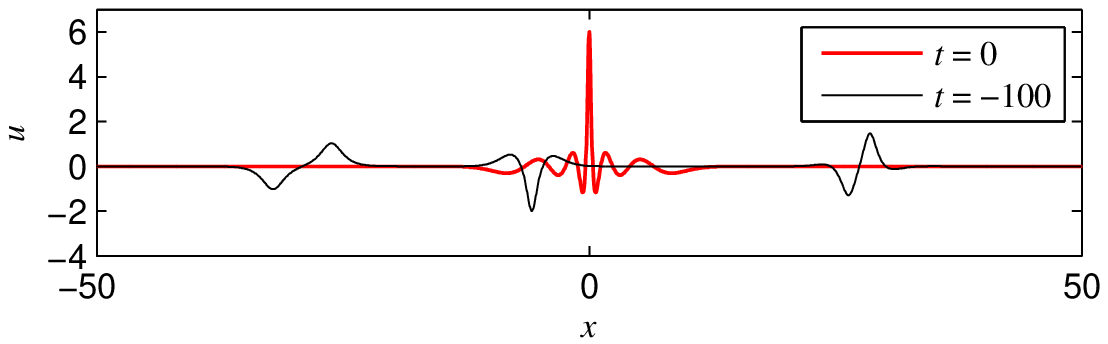}(a)}
\centerline{\includegraphics[width=7.5cm]{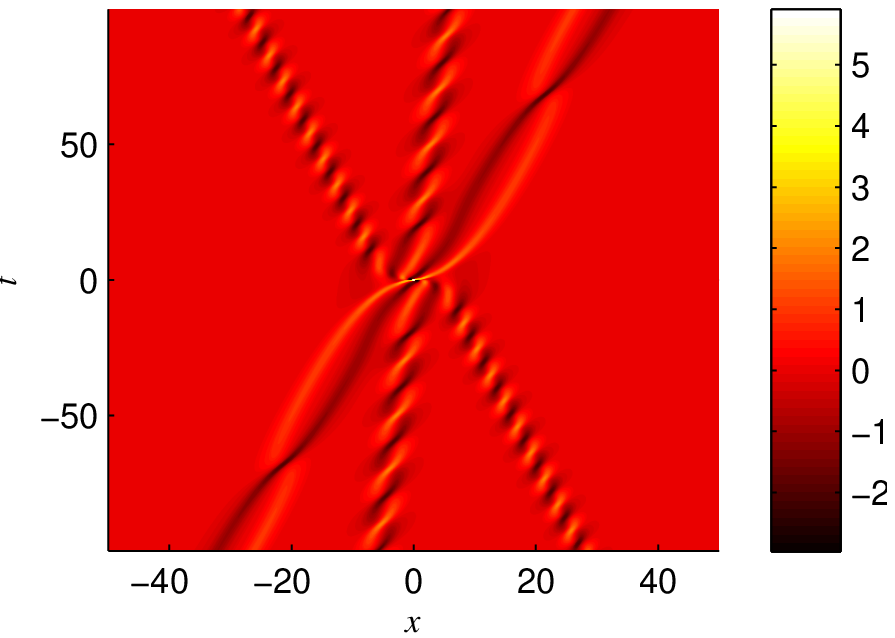}(b)}
\centerline{\includegraphics[width=7.5cm]{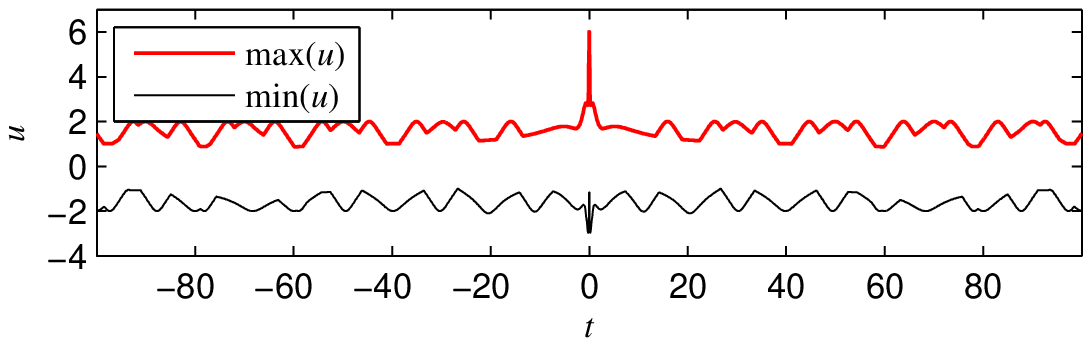}(c)}
\caption{Focusing of three mKdV breathers: the solution before and at the moment of focusing (a), the spatio-temporal plot of $u(x,t)$ (b), and the corresponding temporal evolution of maximum and minimum (c). The breather parameters are $a + ib = \{1+0.8i, 1+0.5i, 1+0.1i\}$.}
\label{fig:3}
\end{figure}

\emph{Rogue waves from collisions of mKdV breathers.}
%
MKdV breathers are specified by quartets of complex conjugated eigenvalues $\lambda  = \{ \pm a/2  \pm ib/2 \}$, where $a$ and $b$ are real values.
A breather solution may be written as
\begin{align}\label{mKdVBreather}
u_{br} = 2ab \frac{a \sinh{\Psi} \sin{\Phi} - b \cosh{\Psi} \cos{\Phi}}
{a^2 \sin^2{\Phi} - b^2 \cosh^2{\Psi}}, \\ \nonumber
\Psi = a \left( x-x_0 - (a^2-3b^2)t \right), \\ \nonumber
\Phi = b \left( x-x_0 - (3a^2-b^2)t \right) + \theta_0 \,.
\end{align}
where $\Psi$  controls the wave envelope,
and  $\Phi$ corresponds to the inner wave.
 In case $|b|  \ll |a|$ the breather resembles an everlasting collision by turns between two solitons of different polarities (see the leftmost group in Fig.~\ref{fig:3}a and corresponding path in Fig.~\ref{fig:3}b); the breather represents a wave packet (the rightmost group in Fig.~\ref{fig:3}a is an example), when $|b| \gg |a|$. In the course of evolution values of $u_{br}(x, t)$ are confined between  $-2|a|$ and $+2|a|$.


Solitons of opposite polarities with close velocities are known to tend to form bound states due to weak perturbations (e.g., weak dissipation), what is described by a bifurcation of two close real eigenvalues to two complex conjugated values \cite{IvanychevFraiman1997}.
%
Solution (\ref{NmKdVSoliton}) may be used for producing multi-breather solutions or combined multi-soliton-breather solutions. A breather with parameters $(a_k, b_k)$ may be built in the exact solution $u_N$ when one pair of parameters $\mu_j$, $\mu_{j+1}$ is specified in the form
\begin{align}\label{BreatherParameters}
\mu_j = (a_k+ib_k)/2, \quad \mu_{j+1} = (a_k-ib_k)/2  \,.
\end{align}
Since (\ref{FocusMaximum}) remains valid for complex eigenvalues, a proper choice of breather parameters may provide superposition of the breathers (or breathers and solitons) in phase.

	Non-degenerative multi-breather solutions may consist, for example, of packets with the same amplitudes but different velocities, what is the optimal choice with respect to the maximum amplification $AI$.
Such an example is shown in Fig.~\ref{fig:3}, where solution (\ref{NmKdVSoliton}) describes the evolution of three breathers with equal amplitudes.
 They collide and lead to  formation of a transient wave with thrice larger amplitude.
%
%
The maximum amplitude of $u_N$ before the collision oscillates and does not exceed $2|a|$ (Fig.~\ref{fig:3}c). Fig.~\ref{fig:3}b displays how faster breathers outrun the slower ones.
 The behaviour of the solution close to the focusing point (Fig.~\ref{fig:3}b) is very similar to the case of multi-soliton interaction (Fig.~\ref{fig:2}b, note different scales in the figures), breathers delay oscillating for some time, the rogue event is characterized by a very short lifetime.


%

\emph{Conclusion.}
%
The effect of multiple soliton interactions on properties of soliton ensembles strongly depends on details of the collision process.
In the case of KdV-type equations for real-valued fields solitons own 'frozen' phases (polarities). Unipolar solitons repulse, and thus do not born higher waves at all. In contrast, solitons of opposite polarities which may coexist within the mKdV framework can virtually inverse the phase of the slower soliton, and then much higher waves occur. This process is not restricted to pairs of bipolar solitons, but may involve unlimited number of solitons resulting in nontrivial dynamics of the solution. This matter requires involved statistical description beyond the paradigm of pairwise soliton interactions, which does not exist at present.

A simultaneous intersection of soliton trajectories is necessary but not sufficient for the efficient focusing of soliton trains. When the solitons approach the focusing point, they are positioned in the order of descending velocities. In addition they should have alternate polarities, what provides the most optimal pattern for generation of extreme bursts. Then the wave amplitude in focus is just the sum of heights of the focusing solitons. Thus the maximum wave amplification is limited only by the number of interacting solitons. Trains of mKdV breathers, which may be considered as coupled solitons of opposite polarity, behave similarly. They also may be targeted to focus in phase; then the partial breather amplitudes sum up.

The presented approach for constructing soliton and breather trains which cause rogue events is not confined to the framework of the modified KdV equation. We have checked that similar scenarios take place in other integrable systems which admit soliton solutions with more than one allowed phase/polarity, such as focusing Gardner equation (quadratic-cubic extension of the KdV equation) and focusing nonlinear Schr\"{o}dinger equation. These results will be reported in a following publication.

The analytic study was supported by the RSF grant No 16-17-00041. The complementary numerical simulations were performed within the Volkswagen Foundation project and RFBR grants 14-02-00983 and 16-55-52019.

\end{document}